\documentclass[12pt]{article}

\usepackage{epsfig}

\newcommand{\be}{\begin{eqnarray}}
\newcommand{\ee}{\end{eqnarray}}

\def\lsim{\mathrel{\rlap{\lower4pt\hbox{\hskip1pt$\sim$}}
    \raise1pt\hbox{$<$}}}               
\def\gsim{\mathrel{\rlap{\lower4pt\hbox{\hskip1pt$\sim$}}
    \raise1pt\hbox{$>$}}}               

\begin{document}

\rightline{}

\begin{center}

\LARGE{Possible evidence of extended objects\\[2mm] inside the proton}

\vspace{1cm}

\large{R. Petronzio$^1$, S. Simula$^2$ and G. Ricco$^3$}

\vspace{0.5cm}

\normalsize{$^1$Dip. di Fisica, Universit\`a di Roma "Tor Vergata" and INFN, Sezione di Roma II,\\ Via della Ricerca Scientifica 1, I-00133 Roma, Italy\\$^2$Istituto Nazionale di Fisica Nucleare, Sezione di Roma III,\\ Via della Vasca Navale 84, I-00146, Roma, Italy\\$^3$Dip. di Fisica, Universit\`a di Genova and INFN, Sezione di Genova,\\ Via Dodecanneso 33, I-16146, Genova, Italy}

\end{center}

\vspace{1cm}

\begin{abstract}

\noindent Recent experimental determinations of the Nachtmann moments of the inelastic structure function of the proton $F_2^p(x, Q^2)$, obtained at Jefferson Lab, are analyzed for values of the squared four-momentum transfer $Q^2$ ranging from $\approx 0.1$ to $\approx 2 ~ (GeV/c)^2$. It is shown that such inelastic proton data exhibit a new type of scaling behavior and that the resulting scaling function can be interpreted as a constituent form factor consistent with the elastic nucleon data. These findings suggest that at low momentum transfer the inclusive proton structure function originates mainly from the elastic coupling with extended objects inside the proton. We obtain a constituent size of $\approx 0.2 \div 0.3 ~ fm$.

\end{abstract}

\vspace{2cm}


\newpage

\pagestyle{plain}

\section{Introduction}

\indent Since long time hadronic spectroscopy and Deep Inelastic Scattering ($DIS$) data have been the two main sources of information on hadron structure. The investigation of hadron mass spectra has led to the introduction of the concept of quarks \cite{quark}, leading to the very fruitful idea that meson and baryons are bound-states of two and three quarks. Such quarks are commonly referred to as Constituent Quarks ($CQ$'s). The $DIS$ data (starting from the pioneering experiments at $SLAC$ in the sixties \cite{SLAC}) have been successfully interpreted in terms of a short-distance partonic structure of the hadrons, i.e. the presence of point-like constituents inside the hadrons \cite{parton}.

\indent With the advent of Quantum ChromoDynamics ($QCD$) partons have been identified with current quarks and gluons, i.e. with the fundamental degrees of freedom of the $QCD$ Lagrangian. On the contrary, a rigorous derivation of the $CQ$'s from $QCD$ is lacking, but $CQ$'s are commonly believed to be quasi-particles emerging from the dressing of valence quarks with gluons and quark-antiquark pairs. If $CQ$'s are confined objects, they should be connected each other by color strings, which may have their own partonic content. In the resolution range in which the sea-quark and gluon content of the strings is not probed, one is naturally lead to try to explain the $DIS$ data only in terms of $CQ$'s having a structure.

\indent The idea to use $CQ$'s as an intermediate step between the current quarks and the hadrons is not new at all and indeed it dates back to the seventies \cite{Altarelli}. There a two-stage model for the parton distributions was proposed, in which any hadron contains a finite number of $CQ$'s having a partonic structure. The latter depends only on short-distance (high-$Q^2$) physics, which is independent of the particular hadron, while the motion of the $CQ$'s inside the hadron reflects the non-perturbative (low-$Q^2$) physics, which depends on the particular hadron. Therefore, within such a picture the $DIS$ structure function of a hadron $F_2^H(x, Q^2)$ can be simply written as the convolution of the structure function of the constituents $F_2^j(x/z, Q^2)$ with the light-front ($LF$) momentum distribution $f_j^H(z)$ of the $j$-th constituent inside the hadron $H$, viz.
 \be
    F_2^H(x, Q^2) = \sum_j ~ \int_x^1 dz ~ f_j^H(z) ~ F_2^j({x \over z}, 
    Q^2)
    \label{eq:convolution}
 \ee
where $z$ is the $LF$ momentum fraction carried by the constituent in the hadron. A convolution analogous to Eq.~(\ref{eq:convolution}) holds as well for each partonic density in the hadron in terms of the corresponding partonic density inside the constituents. The latter can be obtained by a deconvolution of available data on a hadron $H$, provided a reasonable model for the wave function describing the motion of the constituents in the hadron $H$ is considered. Then the structure function of a different hadron $H'$ can be predicted once its wave function is given. Such a procedure has been applied in Ref.~\cite{APR} to predict the structure function of the pion from the known nucleon structure function, and the final result was that the two-stage model based on Eq.~(\ref{eq:convolution}) is supported by data, at least as a first good approximation.

\indent The following question naturally arises: is the two-stage model a good approximation also far from the deep inelastic regime ? In particular, can the model be generalized in such a way to predict hadron structure functions for values of $Q^2$ below and around the scale of chiral symmetry breaking, $\Lambda_{\chi} \approx 1 ~ GeV$ ? The aim of this paper is to answer such a question by extending the two-stage model in order to include the low-$Q^2$ regime and to test it against recent proton structure function data obtained in Hall B at Jefferson Lab with the $CLAS$ spectrometer \cite{JLAB}. It will be shown that the data exhibit a new type of scaling behavior, expected within the generalized two-stage model, and that the resulting scaling function can be interpreted as a $CQ$ form factor consistent with the elastic proton (and neutron) data. These findings suggest that at low momentum transfer the inclusive proton structure function originates mainly from the elastic coupling with {\em extended objects inside the proton}. We obtain a $CQ$ size of $\approx 0.2 \div 0.3 ~ fm$.

\indent The plan of the paper is as follows. The generalization of the original two-stage model to low values of $Q^2$ is presented in Section II and a new type of scaling behavior, which should hold for the moments of the structure function, is proposed. In Section III the basic theoretical input quantity, i.e. the $LF$ momentum distribution $f_j^H(z)$ of a $CQ$ inside the hadron, is discussed and estimated in case of the proton. In Section IV we investigate the possible occurrence of the new scaling property in the recent $JLab$ data \cite{JLAB}, as well as the possible interpretation of the resulting scaling function as the first experimental evidence of the $CQ$ form factor. Our conclusions are summarized in Section V.

\section{Extension of the two-stage model to low momentum transfer}

\indent In this Section the original two-stage model of Refs.~\cite{Altarelli,APR} will be generalized in order to include the low-$Q^2$ regime. As a first step, let us develop such a generalization in a simplified form, which avoids many complications in the final formulae arising from a complete treatment of finite-$Q^2$ effects, but at the same time illustrates the essential physical motivations. The proper treatment of kinematical finite-$Q^2$ effects will be recovered later on in Section IV.

\indent In a $DIS$ experiment at high values of $Q^2$ the internal structure of a $CQ$ is probed, whereas for sufficiently low values of $Q^2$ such a structure cannot be resolved any more. Generally speaking, we expect that the turning point between the high-$Q^2$ and low-$Q^2$ regimes is around the scale of chiral symmetry breaking, $\Lambda_{\chi} \approx 1 ~ GeV$. As $Q^2$ decreases below $\approx \Lambda_{\chi}^2$, we have two expectations: ~ i) the {\em inelastic} coupling of the incoming virtual boson with the $CQ$ becomes less and less important, at least because final states are limited by phase space effects; ~ ii) the {\em elastic} coupling of the incoming virtual boson with the $CQ$ becomes more and more important. We point out that at very low values of $Q^2$ of the order of $\Lambda_{QCD}^2$ [$\approx 0.1 \div 0.2 ~ (GeV/c)^2$] the reinteractions among $CQ$'s in the final state, which are not considered in our present analysis, cannot be neglected any more (see later on Section~3). Therefore, the $Q^2$-range where we want to extend the two-stage model is qualitatively given by $0.1 \div 0.2 \lsim Q^2 ~ (GeV/c)^2 \lsim 1 \div 2$.  

\indent Let us start by writing the $CQ$ structure function $F_2^j$ appearing in the convolution formula (\ref{eq:convolution}) as the sum of two terms $F_2^j = F_2^{j (inel)} + F_2^{j (el)}$, corresponding respectively to the inelastic and elastic virtual boson coupling with the $CQ$. Then, the inelastic structure function of a hadron, $F_2^H(x, Q^2)$, can be written as the sum of two terms, representing the inelastic and elastic $CQ$ contributions, respectively. One has
 \be
    F_2^H(x, Q^2) = \sum_j ~ \int_x^1 dz ~ f_j^H(z) ~ F_2^{j (inel)}({x 
    \over z}, Q^2) + \sum_j ~ \int_x^1 dz ~ f_j^H(z) ~F_2^{j (el)}({x \over 
    z}, Q^2) 
    \label{eq:sum}
 \ee
where, as previously anticipated, we have kept the simplified convolution form in order to avoid up-to-now inessential complications due to finite $Q^2$. The elastic part of the $CQ$ structure function reads explicitly as
 \be
    F_2^{j (el)}(x', Q^2) = G_j^2(Q^2) \delta(x' - 1)
    \label{eq:CQel}
 \ee
where
 \be
    [G_j(Q^2)]^2 = { [G_E^j(Q^2)]^2 + \tau [G_M^j(Q^2)]^2 \over 1 + \tau } 
    = [F_1^j(Q^2)]^2 + \tau [F_2^j(Q^2)]^2
    \label{eq:G2j}
 \ee
with $F_{1(2)}(Q^2)$ and $G_{E(M)}(Q^2)$ representing the Dirac(Pauli) and electric(magnetic) Sachs form factors of the $j$-th $CQ$, respectively. Finally, in Eq.~(\ref{eq:G2j}) $\tau \equiv Q^2 / 4 m_j^2$ with $m_j$ being the $j$-th $CQ$ mass. Thus, the inelastic structure function of the hadron $H$ becomes
 \be
    F_2^H(x, Q^2) = \sum_j ~ \int_x^1 dz ~ f_j^H(z) ~ F_2^{j (inel)}({x 
    \over z}, Q^2) + \sum_j [G_j(Q^2)]^2 ~ x \cdot f_j^H(x)
    \label{eq:inel+el}
 \ee
In the $DIS$ regime the elastic $CQ$ contribution is suppressed by the $CQ$ form factors and one gets
 \be
    F_2^H(x, Q^2) \to_{DIS} \sum_j ~ \int_x^1 dz ~ f_j^H(z) ~ F_2^{j 
    (inel)}({x \over z}, Q^2)
    \label{eq:highQ2}
 \ee
On the contrary, for low values of $Q^2$ the inelastic $CQ$ contribution is expected to become negligible and one could have
 \be
    F_2^H(x, Q^2) \to_{\Lambda_{QCD}^2 \lsim Q^2 \lsim \Lambda_{\chi}^2} 
    \sum_j [G_j(Q^2)]^2 ~ x \cdot f_j^H(x)
    \label{eq:lowQ2}
 \ee

\indent However, it should be immediately realized that Eq.~(\ref{eq:lowQ2}) cannot hold at each $x$ value. Indeed, at low $Q^2$ the hadron structure function $F_2^H(x, Q^2)$ is characterized by resonance bumps emerging over a smooth background, whereas the elastic $CQ$ contribution is expected to have a smooth $x$-shape only, governed by the $LF$ momentum distributions $f_j^H(x)$. Therefore, we assume that Eq.~(\ref{eq:lowQ2}) holds in a dual sense: the $x$-averages of $F_2^H$ over each of the resonance bumps are representative of the elastic $CQ$ contribution [see the r.h.s. side of Eq.~(\ref{eq:lowQ2})] at the corresponding mean values of $x$. Such a {\em $CQ$-hadron duality} can be conveniently expressed in terms of moments of the hadron structure function, defined as
 \be
    M_n^H(Q^2) \equiv \int_0^1 dx ~ x^{n-2} ~ F_2^H(x, Q^2)
    \label{eq:CN}
 \ee
In a similar way we can define the {\em dual} moments as the moments of the elastic $CQ$ contribution, given by
 \be
    M_n^{dual}(Q^2) = \int_0^1 dx ~ x^{n-2} ~ \sum_j [G_j(Q^2)]^2 ~ x \cdot 
    f_j^H(x)
    \label{eq:dualCN}
  \ee
The occurrence of a $CQ$-hadron duality for $Q^2 \lsim \Lambda_{\chi}^2$ can be now translated into the dominance of the dual moments $M_n^{dual}(Q^2)$ for low values of $n$, viz.
 \be
    M_n^H(Q^2) \simeq M_n^{dual}(Q^2)
    \label{eq:CQdual}
 \ee
The limitation to low values of $n$ arises from the fact that as $n$ increases the moment $M_n^H(Q^2)$ is more and more sensitive to the rapidly varying bumps of the resonances. Therefore Eq.~(\ref{eq:CQdual}) cannot hold at very large values of $n$ (see Refs.~\cite{RGP,Ricco,SIM00} for the case of the parton-hadron Bloom-Gilman duality \cite{BG}). At the same time it should be pointed out that the dual relation (\ref{eq:CQdual}) is expected to hold only for $n > 2$, because the second moment $M_2(Q^2) = \int_0^1 dx ~ F_2^H(x, Q^2)$ is significantly affected by the low-$x$ region where the concept of valence dominance may become unreliable.

\indent Let us introduce the squared form factor $[F(Q^2)]^2$ defined as
 \be
     [F(Q^2)]^2 \equiv { \sum_j [G_j(Q^2)]^2 \over \sum_j e_j^2} = {\sum_j 
     [F_1^j(Q^2)]^2 + \tau [F_2^j(Q^2)]^2 \over \sum_j e_j^2}
     \label{eq:FCQ}
 \ee
which is normalized to $1$ at the photon point. Assuming $SU(2)$-symmetric $CQ$ form factors, Eq.~(\ref{eq:dualCN}) becomes
 \be
     M_n^{dual}(Q^2) = [F(Q^2)]^2 \cdot \overline{M}_n^H
     \label{eq:Mndual}
  \ee
with
 \be
     \overline{M}_n^H = \int_0^1 dx ~ x^{n-1} ~ \sum_j ~ e_j^2 ~ f_j^H(x)
     \label{eq:Mbar}
 \ee
If one possesses a reasonable model for the $CQ$ momentum distributions $f_j^H(x)$, the moments $\overline{M}_n^H$ can be estimated and therefore the following ratio
 \be
    R_n^H(Q^2) \equiv M_n^H(Q^2) ~ / ~ \overline{M}_n^H
    \label{eq:ratio}
 \ee
can be constructed starting from the full moments $M_n^H(Q^2)$ [Eq.~(\ref{eq:CN})]. The ratio $R_n^H(Q^2)$ should generally depend on both $n$ and $Q^2$ as well as on the hadron $H$. However, when the underlying $CQ$ picture holds, the $CQ$-hadron duality (\ref{eq:CQdual}) is expected to hold as well and, consequently, the ratio $R_n^H(Q^2)$ depends only on $Q^2$, i.e. it becomes independent of both the order $n$ and the hadron, viz. 
 \be
    R_n^H(Q^2) \simeq [F(Q^2)]^2
    \label{eq:scaling}
 \ee
The scaling function, given by the r.h.s. of Eq.~(\ref{eq:scaling}), is directly the square of the $CQ$ form factor, i.e. the form factor of a confined object. The important point is that within our generalized two-stage model the new scaling property (\ref{eq:scaling}) is expected to occur at low $Q^2$. We point out that, once the $CQ$ form factor is extracted from known hadron data, the moments of the structure function of another hadron can in principle be predicted.

\indent Let us now introduce the recent results obtained at $JLab$ \cite{JLAB}, where the inclusive electron-proton cross section has been measured in the nucleon resonance regions for values of $Q^2$ below $4.5 ~ (GeV/c)^2$ using the $CLAS$ detector. One of the relevant feature of such measurements is that the $CLAS$ large acceptance has allowed to determine the cross section in a wide two-dimensional range of values of $Q^2$ and $x$ and has made it possible to directly integrate all the existing data at fixed $Q^2$ over the whole significant $x$-range for the determination of the proton moments $M_n^p(Q^2)$ with order $n \geq 2$. More precisely, the Nachtmann proton moments, defined as \cite{Nachtmann}
 \be
    M_n^p(Q^2) \equiv \int_0^1 dx {\xi^{n+1} \over x^3} {3 + 3 (n + 1) r + n 
    (n + 2) r^2 \over (n + 2) (n + 3)} ~ F_2^p(x,Q^2)
    \label{eq:Mnp}
 \ee
where $r \equiv \sqrt{1 + 4 M^2 x^2 / Q^2}$ and $\xi \equiv 2 / (1 + r)$, have been directly extracted from the data for $n = 2, 4, 6, 8$ \cite{JLAB}. As it is well known, the main advantage of the Nachtmann moments (\ref{eq:Mnp}) over the Cornwall-Norton moments (\ref{eq:CN}) is that only with the former it is possible to cancel out all the finite-$Q^2$ kinematical corrections due to the non-vanishing mass of the target. Thus, in what follows Eq.~(\ref{eq:Mnp}) replaces Eq.~(\ref{eq:CN}) for $H = p$.

\begin{figure}[htb]

\centerline{\epsfig{file=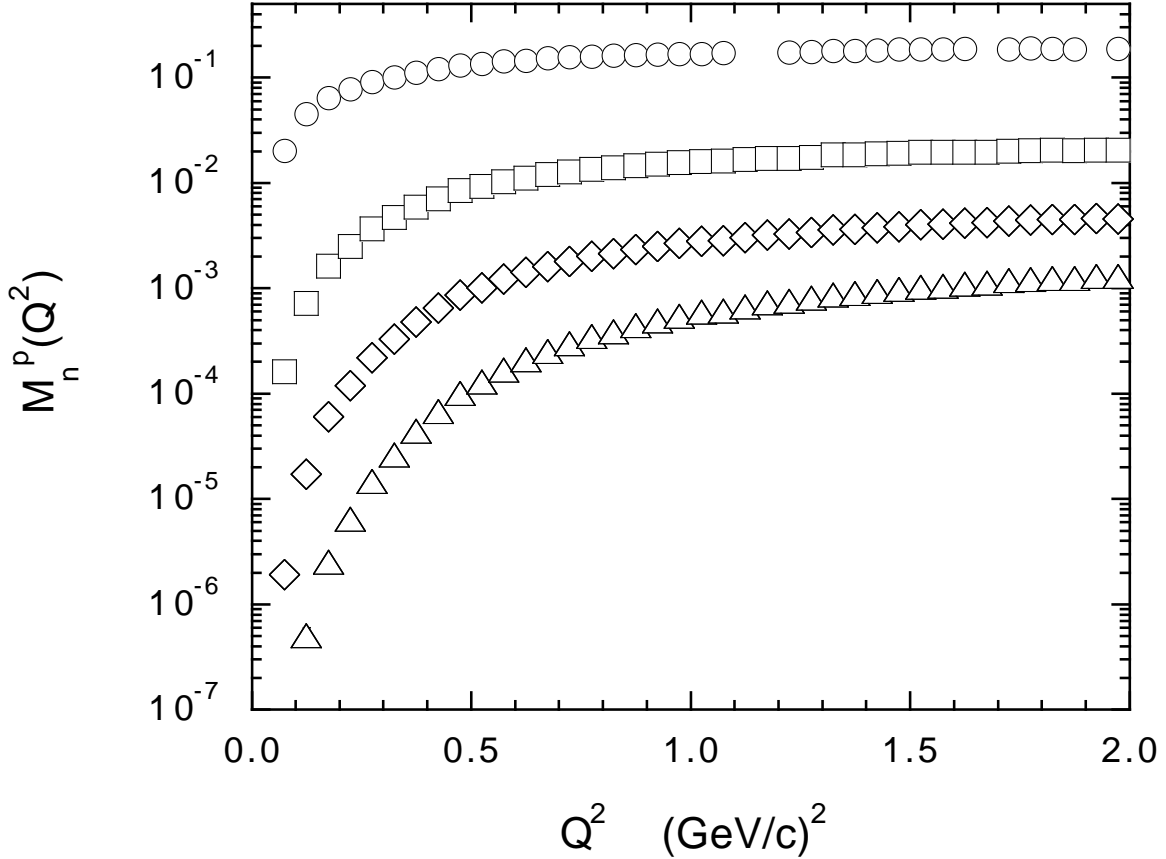,width=15.5cm}}

\caption{\label{fig:Mnp} \small \em Experimental (inelastic) Nachtmann moments $M_n^p(Q^2)$ of the proton versus $Q^2$ from Ref.~\cite{JLAB}. The dots, squares, diamonds and triangles correspond to $n = 2, 4, 6$ and $8$, respectively. The statistical errors are reported, but they are not visible.}

\end{figure}

\indent In Fig.~\ref{fig:Mnp} the {\em experimental} Nachtmann moments $M_n^p(Q^2)$, determined in Ref.~\cite{JLAB}, are shown in the $Q^2$-range of interest for this work, namely $0.1 \lsim Q^2 \lsim 2 ~ (GeV/c)^2$. The contribution arising from the elastic proton peak ($x = 1$) is not included and therefore, from now on, the moments $M_n^p(Q^2)$ represent the inelastic part of the proton Nachtmann moments.

\begin{figure}[htb]

\centerline{\epsfig{file=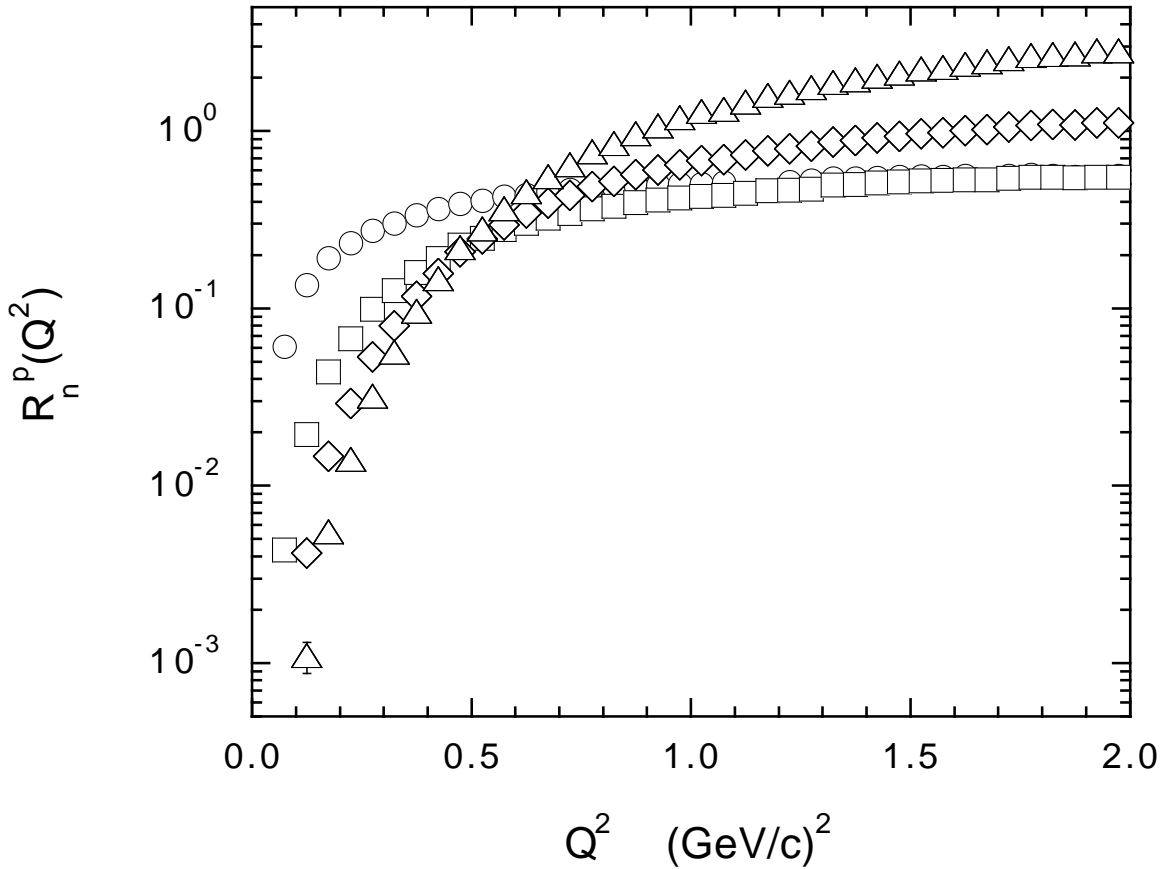,width=15.5cm}}

\caption{\label{fig:delta} \small \em Ratio $R_n^p(Q^2)$ [Eq.~(\ref{eq:ratio}) for $H = p$] calculated using the experimental Nachtmann moments $M_n^p(Q^2)$ [Eq.~(\ref{eq:Mnp})], shown in Fig.~\ref{fig:Mnp}, and assuming a delta-like shape for the $CQ$ momentum distribution in the proton, namely $\sum_j e_j^2 f_j^p(x) = \delta(x - 1/3)$ [see Eq.~(\ref{eq:deltafz})]. The meaning of the markers is the same as in Fig.~\ref{fig:Mnp}.}

\end{figure}

\indent The $Q^2$-behavior of the moments $M_n^p(Q^2)$ shown in Fig.~\ref{fig:Mnp} is characterized by a sharp rise at low $Q^2$, followed by a smoother behavior for $Q^2 \gsim 1 ~ (GeV/c)^2$. However, the dependence upon the order $n$ is much more interesting. Indeed, the moments $M_n^p(Q^2)$ appear to differ by approximately an order of magnitude moving from $n$ to $n + 2$. As a result, though the range of values considered for $n$ is quite restricted ($2 \leq n \leq 8$), the values of the corresponding moments are spread over several order of magnitudes. Such a behavior can be qualitatively explained within our generalized two-stage model in the following way. Let us assume a very simplified and quite rough model for the $CQ$ momentum distribution $f_j^p(z)$ in the proton, in which the constituents share exactly just a fraction $1 / 3$ of the proton momentum, viz.
 \be
    \sum_j e_j^2 f_j^p(x) \to \delta(x - 1 /3)
    \label{eq:deltafz}
 \ee
The moments (\ref{eq:Mbar}) simply become
 \be
    \overline{M}_n^p \to \left( {1 \over 3} \right)^{n - 1}
    \label{eq:delta}
 \ee
implying a factor of $\approx 1 / 9$ between the orders $n$ and $n + 2$.
Thus, in Fig.~\ref{fig:delta} we have reported the ratio (\ref{eq:ratio}) obtained using the experimental Nachtmann moments $M_n^p(Q^2)$ [Eq.~(\ref{eq:Mnp})], shown in Fig.~\ref{fig:Mnp}, and assuming Eq.~(\ref{eq:delta}). It can clearly be seen that with respect to the experimental moments $M_n^p(Q^2)$ the spread of the ratio $R_n^p(Q^2)$ as a function of $n$ has been largely reduced. This is an important result obtained with a very simple hypothesis about the $CQ$ motion in the proton. Figure~\ref{fig:delta} shows that there is a clear {\em tendency of the data} toward a {\em scaling property} like Eq.~(\ref{eq:scaling}).

\indent Any way, we have to consider that Eqs.~(\ref{eq:deltafz}-\ref{eq:delta}) imply that the relative motion of the $CQ$'s inside the proton is neglected, which is not a reliable assumption in case of light constituents. Therefore, in the next Section we perform more realistic estimates of the $CQ$ momentum distribution in the proton with the aim of approaching better the scaling property (\ref{eq:scaling}) as well as of interpreting the scaling function as a (squared) form factor.

\section{$CQ$ light-front momentum distributions in the proton}

\indent Within the two-stage model the basic theoretical input quantity, appearing in Eq.~(\ref{eq:Mbar}), is the $LF$ momentum distribution 
 \be
    \overline{f}^H(z) \equiv \sum_j e_j^2 f_j^H(z)
    \label{eq:fbar}
 \ee
Such a distribution results from the motion of the $CQ$'s inside the particular hadron $H$ and in what follows we will explicitly limit ourselves to the case of the proton, which is of interest in this work.

\indent In order to evaluate the constituent $U$ and $D$ quark distributions in the proton it is natural to adopt the Hamiltonian $LF$ formalism \cite{LF}. In terms of the intrinsic $LF$ variables $\xi_i$ and $\vec{k}_{i \perp}$ (see the Appendix for their definition) the $CQ$ momentum distribution in the proton is given by
 \be
    f_Q^p(z) = {3 \over 2} \sum_{\nu_p} \int [d\xi_i d\vec{k}_{i \perp}] 
    \sum_{\{\nu_i \tau_i \}}  ~ \delta(z - \xi_1) ~ \delta_{\tau_Q, \tau_1} 
    ~ \left| \langle \{ \xi_i \vec{k}_{i \perp}; \nu_i \tau_i \} | 
    \Psi_p^{\nu_p} \rangle \right|^2
    \label{eq:fQz}
 \ee
where $Q = U, D$, $\tau_U = 1/2$, $\tau_D = -1/2$ and $[d\xi_i d\vec{k}_{i \perp}]$ stands for $d\vec{k}_{1 \perp} d\vec{k}_{2 \perp} d\vec{k}_{3 \perp} ~ \delta(\vec{k}_{1 \perp} + \vec{k}_{2 \perp} + \vec{k}_{3 \perp}) ~ d\xi_1 d\xi_2 d\xi_3 ~ \delta(\xi_1 + \xi_2 + \xi_3 - 1)$. In Eq.~(\ref{eq:fQz}) $\Psi_p^{\nu_p}$ is the proton $LF$ wave function, whose general structure is briefly illustrated in the Appendix, where also all the other relevant quantities are defined. Note that the $CQ$ distributions (\ref{eq:fQz}) are normalized as
 \be
     \int_0^1 dz ~ f_U^p(z) = 2 ~~~~, ~~~~ \int_0^1 dz ~ f_D^p(z) = 1 ~,
     \label{eq:norm}
 \ee
and satisfy the momentum sum rule
 \be
    \int_0^1 dz ~ z ~ [f_U^p(z) + f_D(z)] = 1 ~;
    \label{eq:MSR}
 \ee
thus, one has $\overline{f}^p(z) = [4 f_U^p(z) + f_D^p(z)] / 9$.
 
\begin{figure}[htb]

\centerline{\epsfig{file=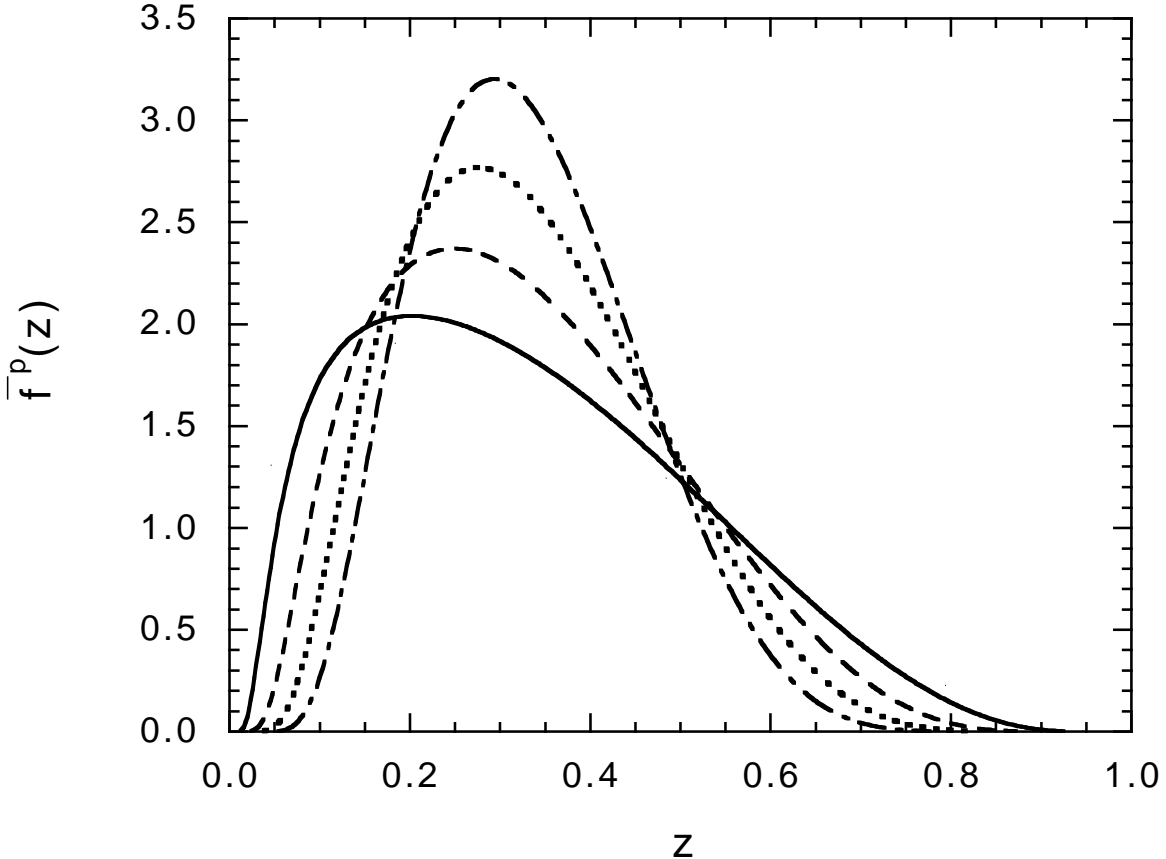,width=15.5cm}}

\caption{\label{fig:fz} \small \em Light-front momentum distribution $\overline{f}^p(z)$ [Eq.~(\ref{eq:SU6})], calculated assuming the $SU(6)$-symmetric gaussian ans\"atz (\ref{eq:gaussian}) for the proton wave function with $\beta = 0.3 ~ GeV$. The solid, dashed, dotted and dot-dashed lines correspond to a $CQ$ mass equal to $m = 0.22, 0.33, 0.44, 0.55 ~ GeV$, respectively.}

\end{figure}

\indent In the Appendix the $CQ$ momentum distributions (\ref{eq:fQz}) are explicitly written in terms of various $SU(6)$ components characterizing the nucleon wave function [see Eq.~(\ref{eq:fUfD})]. If a completely $SU(6)$ symmetric nucleon wave function is considered, one has always $f_U^p(z) = 2 f_D^p(z)$ and therefore the $LF$ momentum distribution $\overline{f}^p(z)$ becomes (cf. the Appendix)
 \be
     \overline{f}^p(z) = \int d\vec{k}_{\perp} d\vec{p}_{\perp} \int 
     [d\xi_i] ~ \delta(z - \xi_1) ~ {E_1 E_2 E_3 \over M_0 \xi_1 \xi_2 
     \xi_3} ~ |w_S(\vec{k}, \vec{p})|^2
     \label{eq:SU6}
 \ee
In order to improve the simple delta-like model given by Eq.~(\ref{eq:deltafz}) we have calculated Eq.~(\ref{eq:SU6}) adopting a gaussian ans\"atz for the proton wave function $w_S(\vec{k}, \vec{p})$, namely
 \be
     w_S(\vec{k}, \vec{p}) \propto e^{- (k^2 + 3 p^2 / 4) / 2 \beta^2}
     \label{eq:gaussian}
 \ee
where $\beta$ is a parameter. The results of our calculations are reported in Fig.~\ref{fig:fz} for various values of the $CQ$ mass $m_U = m_D = m$ keeping the parameter $\beta$ fixed at the value $\beta = 0.3 ~ GeV$, which represents the typical $CQ$ momentum in the proton due to the confinement scale. It can be seen that the calculated distribution $\overline{f}^p(z)$ is peak-shaped with a location of the peak and a width which sharply depend on $m$ for values of $m$ pertaining to the so-called light-$CQ$ sector. The delta-like model (\ref{eq:deltafz}), characterized by a zero-width peak located at $x = 1 / 3$, can be recovered only in the heavy-quark limit $m \to \infty$. As the $CQ$ mass decreases, the width of the peak increases and the location of the peak moves to values of $x$ less than $1 / 3$. Note that: ~ i) the widths are asymmetric around the peaks in order to keep the average fraction of the momentum carried by each $CQ$ equal to $1 / 3$ at any value of $m$, and ~ ii) the distribution $\overline{f}^p(z)$ depends only on the parameter ratio $\beta / m$. Thus, the effects of the $CQ$ motion on the shape of $\overline{f}^p(z)$ are very important and should be taken into account, particularly for light $CQ$ masses.
 
\begin{figure}[htb]

\centerline{\epsfig{file=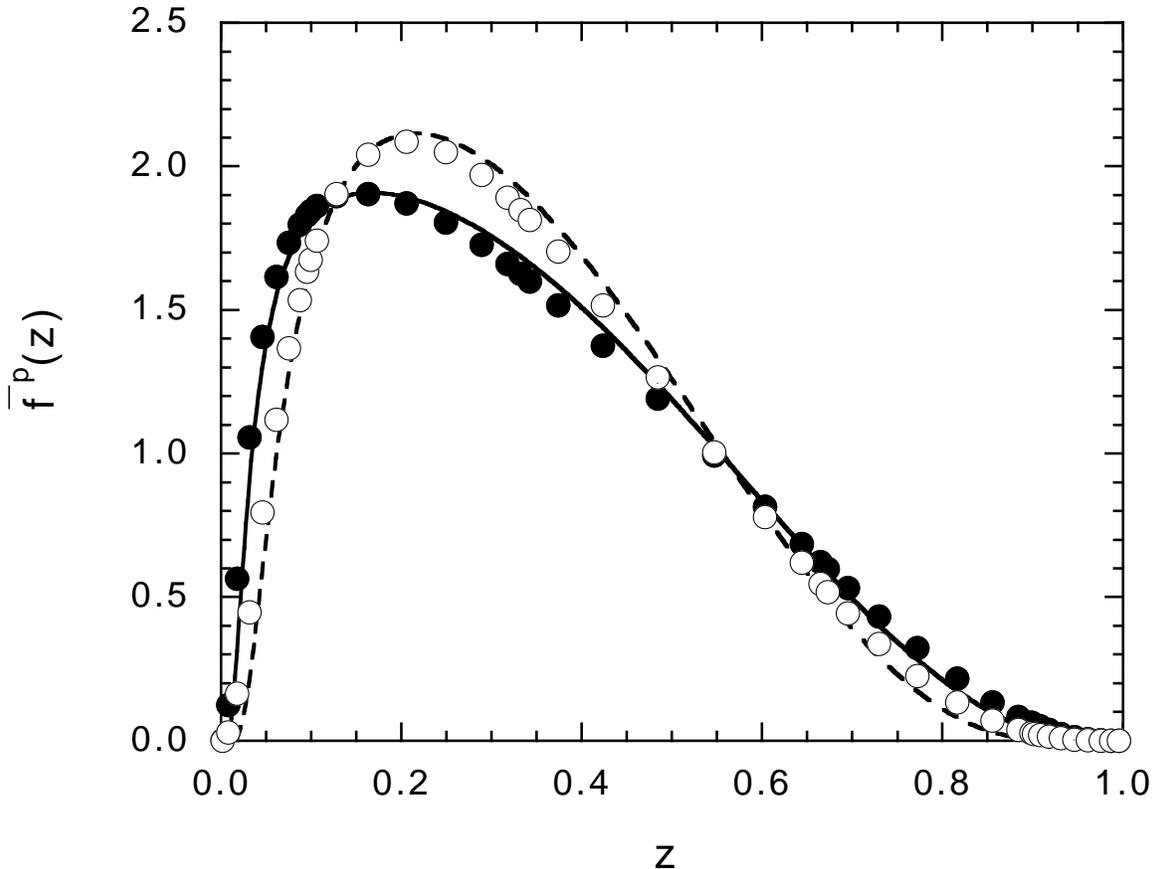,width=15.5cm}}

\caption{\label{fig:compfz} \small \em Light-front momentum distribution $\overline{f}^p(z)$ [Eq.~(\ref{eq:fbar}) for $H = p$], calculated using the full proton wave function corresponding to the one-gluon exchange model of Ref.~\cite{CI} (full dots) and to the chiral model of Ref.~\cite{Glozmann} (open dots). The solid and dashed lines correspond to the case of the  $SU(6)$-symmetric gaussian ans\"atz (\ref{eq:gaussian}) with $\beta / m = 1.2$ and $1.8$, respectively.}

\end{figure}

\indent It is well known (see Ref.~\cite{CI} and references therein) that a good description of hadronic mass spectra requires spin-dependent components in the effective interaction among $CQ$'s. Such components generate $SU(6)$ breakings in the proton wave function (see, e.g., Ref.~\cite{nucleon}). On the contrary the gaussian ans\"atz (\ref{eq:gaussian}) is a pure $SU(6)$-symmetric wave function and therefore we should investigate $SU(6)$-breaking effects in the calculation of the $CQ$ light-front momentum distribution $\overline{f}^p(z)$. To this end we have considered two of the most sophisticated $CQ$ potential models available in the literature, namely the one-gluon-exchange model of Ref.~\cite{CI} and the chiral model of Ref.~\cite{Glozmann}, based on Goldstone-boson-exchange arising from the spontaneous breaking of chiral symmetry. The results obtained for $\overline{f}^p(z)$ are shown in Fig.~\ref{fig:compfz} and compared with those corresponding to the gaussian ans\"atz (\ref{eq:gaussian}) for different values of the parameter ratio $\beta / m$. It can clearly be seen that, as far as $\overline{f}^p(z)$ is concerned, the $SU(6)$ breaking contained in the $CQ$ models of Refs.~\cite{CI,Glozmann} can be approximated to a very good extent by using a gaussian ans\"atz with appropriate values of the parameter ratio $\beta / m$.

\section{Scaling analysis of the experimental moments}

\indent In this Section we apply our generalized two-stage model to the analysis of the data shown in Fig.~\ref{fig:Mnp}, taking into account: ~ i) the motion of the $CQ$'s adopting the gaussian ans\"atz (\ref{eq:gaussian}) for the proton wave function, as described in the previous Section, and ~ ii) the effects of finite $Q^2$, which are expected to be relevant due to the $Q^2$-range of our analysis [$0.1 \div 0.2 \lsim Q^2 ~ (GeV/c)^2 \lsim 1 \div 2$].

\begin{figure}[htb]

\centerline{\epsfig{file=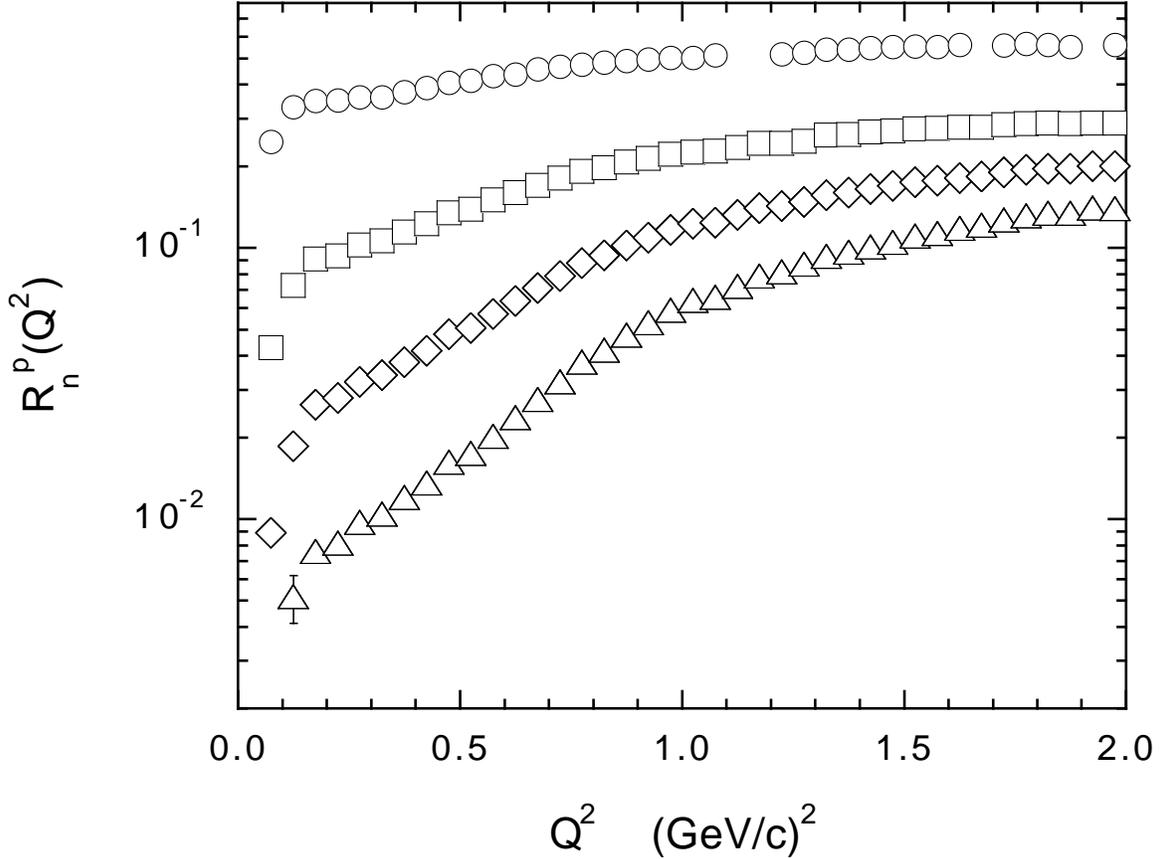,width=15.5cm}}

\caption{\label{fig:gaussian} \small \em Ratio $R_n^p(Q^2)$ [Eq.~(\ref{eq:ratio}) for $H = p$] calculated using the experimental Nachtmann moments $M_n^p(Q^2)$ [Eq.~(\ref{eq:Mnp})] shown in Fig.~\ref{fig:Mnp}, and assuming the gaussian ans\"atz (\ref{eq:gaussian}) for the proton wave function with $\beta = 0.3 ~ GeV$ and $m = 0.25 ~ GeV$. The meaning of the markers is the same as in Fig.~\ref{fig:Mnp}.}

\end{figure}

\indent Let us start by considering the first of the two quoted effects. In Fig.~\ref{fig:gaussian} we have reported the results obtained for the ratio $R_n^p(Q^2)$ calculated using the experimental Nachtmann moments $M_n^p(Q^2)$ [Eq.~(\ref{eq:Mnp})] and assuming the gaussian ans\"atz (\ref{eq:gaussian}) for the proton wave function with $\beta = 0.3 ~ GeV$ and $m = 0.25 ~ GeV$ (corresponding to $\beta / m = 1.2$). The spread of the values of the ratio $R_n^p(Q^2)$ is drastically reduced with respect to the case of the delta-like model (\ref{eq:deltafz}) [cf. Fig.~\ref{fig:delta}]. Note that, as already pointed out (see Section~2) the results at $n = 2$ appear to deviate significantly from those corresponding to larger orders. We have checked that the general qualitative shape of the results shown in Fig.~\ref{fig:gaussian} does not change significantly when the value of the parameter ratio $\beta / m$ is varied.

\indent Though the results shown in Fig.~\ref{fig:gaussian} exhibit a drastic improvement toward a significant reduction in the dependence of the ratio $R_n^p(Q^2)$ upon the order $n$, the scaling property (\ref{eq:scaling}) is still far from being reached. Moreover, the $Q^2$-behavior of $R_n^p(Q^2)$ is completely at variance with what is naturally expected for a squared form factor. The main drawback is clearly the use of Eq.~(\ref{eq:Mbar}), which is meaningful only at large $Q^2$. In our opinion, in order to restore a proper behavior of $R_n^p(Q^2)$, we have to account for "higher-twist" effects, which can be divided into the three following classes:

\begin{itemize}

\item{the inelastic pion threshold, which sets a $Q^2$-dependent maximum value for the $x$-range, given by $x_{max} = x_{\pi} = Q^2 / [Q^2 + (M + m_{\pi})^2 - M^2]$. Note that $x_{\pi}$ largely differs from $1$ at low $Q^2$;}

\item{kinematical power corrections in the physical region $x \leq x_{\pi}$;}

\item{dynamical power corrections due to final state interactions responsible for the resonance bumps in the $x$-space.}

\end{itemize}

\noindent In what follows we will consider the first two effects only. The pion threshold can be simply taken into account by multiplying the distribution $\overline{f}^p(x)$ by a threshold factor $F_{thr}(W)$, where $W$ is the produced invariant mass $W = \sqrt{M^2 + Q^2 (1 - x) / x}$, having the property $F_{thr}(W \leq M + m_{\pi}) = 0$ and $F_{thr}(W \to \infty) = 1$. A simple and parameter-free choice dictated by pure phase space effects is
 \be
    F_{thr}(W) = \sqrt{1 - \left( {M + m_{\pi} \over W} \right)^2}
    \label{eq:threshold}
 \ee
We stress that by means of $F_{thr}(W)$ we account for that part of higher twists which are related to the final-state phase-space constraint.

\indent The kinematical corrections to Eq.~(\ref{eq:Mbar}) originate from the non-vanishing value of the target mass, i.e. the proton mass $M$. The way to construct such corrections is well known in $DIS$ \cite{TM} and therefore, for analogy, we replace the distribution $\overline{f}^p(x)$ by the quantity $\overline{f}_{TM}^p(\xi, Q^2)$, given explicitly by
 \be
    \overline{f}_{TM}^p(\xi, Q^2) & = & {x^2 \over r^3} {\overline{f}^p(\xi) 
    \over \xi^2} + {6 M^2 \over Q^2} {x^3 \over r^4} \int_{\xi}^{\xi^*} 
    d\xi' {\overline{f}^p(\xi') \over \xi' \xi} \nonumber \\
    & + & {12 M^4 \over Q^4} {x^4 \over r^5} \int_{\xi}^{\xi^*} d\xi' 
    {\overline{f}^p(\xi') \over \xi' \xi} (\xi' - \xi) 
    \label{eq:TM}
 \ee
where $\xi$ is the Nachtmann variable, $x = \xi / (1 - M^2 \xi^2 / Q^2)$ and $\xi^* \equiv \mbox{min}(1, Q / M)$ is the maximum allowed value of $\xi$ (cf. Ref.~\cite{SIM00}). It should be reminded that the value $\xi^*$ is larger than the inelastic pion threshold $\xi_{\pi}$. Therefore, the support in which the function $\overline{f}_{TM}^p(\xi, Q^2)$ is defined, contains an unphysical region extending from $\xi_{\pi}$ to $\xi^*$.

\indent We point out that Eq.~(\ref{eq:TM}) expresses the fact that the asymptotic function $\overline{f}^p$ receives a series of power corrections having a scale of order of the proton mass $M$. When the threshold factor $F_{thr}(W)$ is neglected (i.e., $F_{thr}(W) = 1$), the use of the Nachtmann moments cancel out exactly all the power corrections contained in the r.h.s. of Eq.~(\ref{eq:TM}). On the contrary, when the threshold factor is considered (i.e., $F_{thr}(W) \neq 1$), only part of the target-mass corrections can be reabsorbed by the use of the Nachtmann moments. As a matter of facts, for consistency with the experimental data shown in Fig.~\ref{fig:Mnp}, the Cornwall-Norton moment (\ref{eq:Mbar}) has to be replaced by a Nachtmann one. In doing that the quantity $\overline{M}_n^p$ is no more independent of $Q^2$, and therefore Eqs.~(\ref{eq:Mndual}-\ref{eq:Mbar}) are now replaced by
 \be
     M_n^{dual}(Q^2) = [F(Q^2)]^2 \cdot \overline{M}_n^p(Q^2)
     \label{eq:Mndual_Q2}
  \ee
with
 \be
    \overline{M}_n^p(Q^2) & \equiv & \int_0^{\xi^*} d\xi {\xi^{n+1} \over 
    x^3} {3 + 3 (n + 1) r + n (n + 2) r^2 \over (n + 2) (n + 3)} 
    \nonumber \\[2mm] 
    & \cdot & {r (1 + r) \over 2} ~ \xi \overline{f}_{TM}^p(\xi, Q^2) ~ 
    F_{thr}(W) ~~~~~~~~
    \label{eq:Mtheor}
 \ee
where $r (1 + r) / 2 = dx / d\xi$ arises from the change of variables from $x$ to $\xi$. In Eq.~(\ref{eq:Mtheor}) we have put $\xi^*$ as the upper limit of integration; however, due to the threshold factor (\ref{eq:threshold}) the integration extends only up to $\xi_{\pi}$ and therefore part of the target-mass corrections survives after integration. We stress again that this is an important point, because Eq.~(\ref{eq:Mtheor}) reduces exactly to Eq.~(\ref{eq:Mbar}) when the threshold factor $F_{thr}(W)$ is disregarded\footnote{More precisely, when the threshold factor $F_{thr}(W)$ is disregarded, Eq.~(\ref{eq:Mtheor}) reduces at any value of $Q^2$ to $\int_0^{x^*} dx ~ x^{n - 1} \overline{f}^p(x)$, where $x^* = \mbox{min}(1, Q / M)$, as it can be easily checked numerically.}, in agreement with the properties of the Nachtmann moments.

\begin{figure}[htb]

\centerline{\epsfig{file=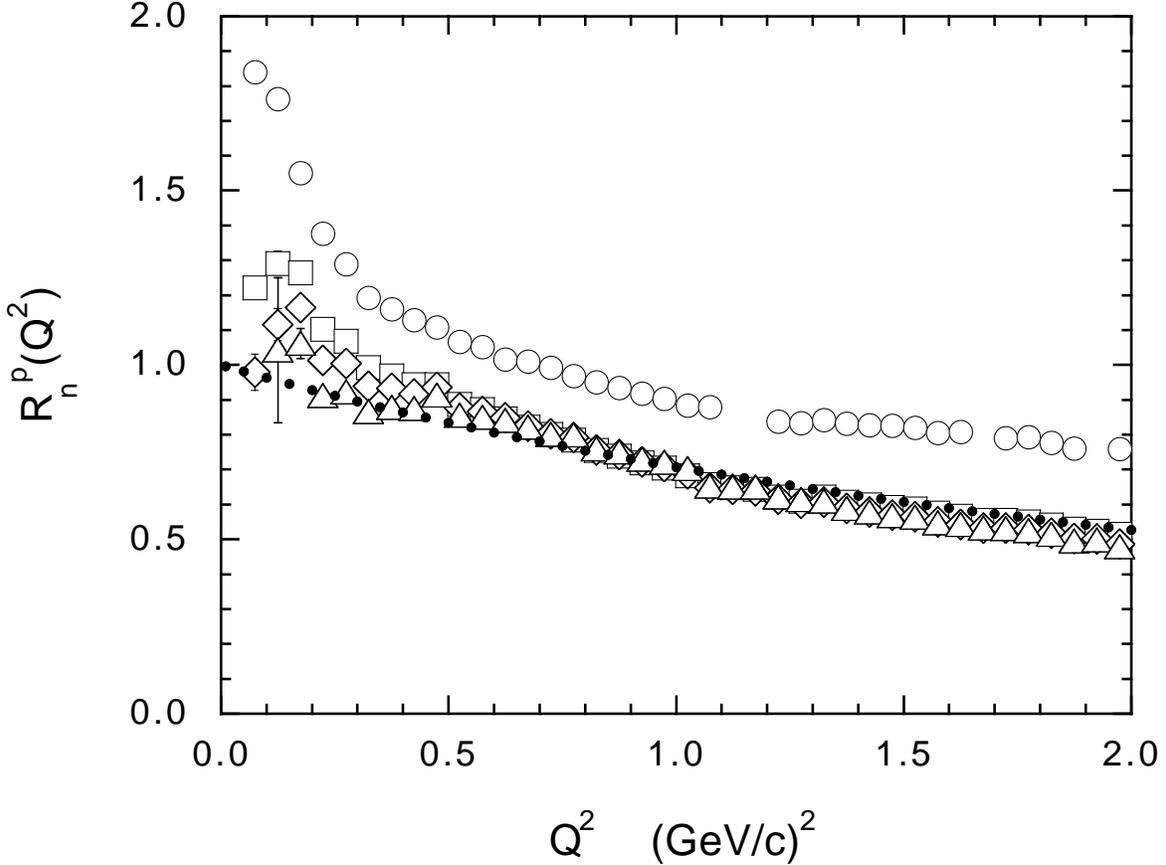,width=15.5cm}}

\caption{\label{fig:threshold} \small \em Ratio $R_n^p(Q^2)$ [Eq.~(\ref{eq:ratio}) for $H = p$] calculated using the experimental Nachtmann moments $M_n^p(Q^2)$ [Eq.~(\ref{eq:Mnp})] shown in Fig.~\ref{fig:Mnp}, and the theoretical moments $\overline{M}_n^p(Q^2)$ given by Eq.~(\ref{eq:Mtheor}). The momentum distribution $\overline{f}^p(\xi)$ corresponds to the gaussian ans\"atz (\ref{eq:gaussian}) for the proton wave function with $\beta = 0.3 ~ GeV$ and $m = 0.25 ~ GeV$. The dotted line represents the squared monopole form factor $[F(Q^2)]^2 = 1 / (1 + r_Q^2 ~ Q^2 / 6)^2$ corresponding to a $CQ$ size equal to $r_Q = 0.21 ~ fm$. The meaning of the markers is the same as in Fig.~\ref{fig:Mnp}.}

\end{figure}

\indent We have calculated Eq.~(\ref{eq:Mtheor}) using the target-mass corrected momentum distribution $\overline{f}^p(\xi, Q^2)$ starting from the gaussian ans\"atz (\ref{eq:gaussian}) for the proton wave function and adopting the threshold factor (\ref{eq:threshold}). The results for the moment ratio $R_n^p(Q^2)$ obtained at $\beta = 0.3 ~ GeV$ and $m = 0.25 ~ GeV$ are reported in Fig.~\ref{fig:threshold}. It can clearly be seen that {\em the scaling property (\ref{eq:scaling}) holds at $n > 2$} even in a linear scale. Moreover, the scaling function closely resembles a squared monopole form factor $[F(Q^2)]^2 = 1 / (1 + r_Q^2 ~ Q^2 / 6)^2$ corresponding to a $CQ$ size equal to $r_Q = 0.21 ~ fm$.

\indent The quality of the scaling exhibited in Fig.~\ref{fig:threshold} is extremely good for $Q^2 \gsim 0.3 ~ (GeV/c)^2$, while it deteriorates at very low values of $Q^2$ (but still the scaling is approximately satisfied within $\approx 30 \%$ even at $Q^2 \approx 0.1 ~ (GeV/c)^2$). This finding is not surprising at all and it can be understood as follows. Let us consider the Operator Product Expansion ($OPE$) of the moments of the proton structure function in terms of local operators acting on elementary (point-like) fields. The so-called higher twists are known to describe correlations among partons. Their contribution to the $OPE$ is given by matrix elements of a series of several operators $O_n$ producing power suppressed terms of the form $(\Lambda_n^2 / Q^2)^{(\tau_n - 2) / 2}$, where $\tau_n$ is the twist and $\Lambda_n$ is the scale associated with the operators $O_n$. The scale $\Lambda_n$ is expected to be proportional to $1 / R_n$, where $R_n$ is the typical average distance of the partonic correlations generated by the operators $O_n$. Which kind of higher twists are accounted for by the spatial extension of the $CQ$'s~? It is clear that we can distinguish two basic types of partonic correlations: those among partons inside the $CQ$ and those between partons belonging to different $CQ$'s, which means correlations between $CQ$'s (in the final state). The former are characterized by a value of $R_n$ close to the $CQ$ size, while the latter correspond to a larger value of $R_n$ of the order of the confinement (hadronic) size. Correspondingly, the scale $\Lambda_n$ is larger for partonic correlations inside the $CQ$ and smaller for partonic correlations among different $CQ$'s. In our model only the first type of higher twists can be thought to be accounted for by the $CQ$ form factor in some effective way\footnote{Indeed, there is no rigorous derivation of the $CQ$ picture from $QCD$.}. Our model does not include power corrections arising from correlations among different $CQ$'s in the final state. Such "long-range" higher twists have a low scale of the order of $\Lambda_{QCD}$, and therefore we expect that they should play an important role mainly for $Q^2 \lsim \Lambda_{QCD}^2 \approx 0.1 \div 0.2 ~ (GeV/c)^2$, i.e. in the $Q^2$-range where the scaling shown in Fig.~\ref{fig:threshold} is only approximate. The estimate of the effects of such "long-range" higher twists is not an easy task and it is well beyond the aim of the present paper. Note that the role of the "long-range" higher twists is even more evident in the $x$-space, because these higher twists are responsible for the huge resonance bumps which are known to characterize the structure function $F_2^p(x, Q^2)$ at low values of $Q^2$. 

\indent We should now investigate the impact of different choices of the functional form of the threshold factor $F_{thr}(W)$ as well as of different values of the parameter ratio $\beta / m$. We have found that the scaling property (\ref{eq:scaling}), clearly exhibited in Fig.~\ref{fig:threshold}, is not very sensitive to the specific choice of $F_{thr}(W)$ and of the parameter ratio $\beta / m$. On the contrary the shape of the scaling function is affected both by the choice of $F_{thr}(W)$ and by the value of the parameter ratio $\beta / m$. It turns out that: ~ i) the use of the specific form (\ref{eq:threshold}) minimizes the scaling violation at the lowest $Q^2$; ~ ii) when the ratio $\beta / m$ changes from the value $1.2$, considered in Fig.~\ref{fig:threshold}, to the value $1.8$, the $CQ$ size $r_Q$ changes correspondingly from $0.21 ~ fm$ to $0.27 ~ fm$.

\indent We point out that an important consistency requirement can be formulated: the $CQ$ form factor extracted from the scaling function and the model used for the wave function should be consistent with elastic nucleon data. This is {\em a crucial requirement} necessary to interpret the scaling function as a (squared) form factor and consequently to get an estimate of the $CQ$ size. To check this point we have calculated the nucleon elastic form factors adopting the covariant $LF$ approach of Ref.~\cite{nucleon}. There the one-body approximation for the electromagnetic ($e.m.$) current operator $J^{\mu}$ is adopted, viz.
 \be
    J^{\mu} \simeq J_1^{\mu} = \sum_j \left[ F_1^j(Q^2) \gamma^{\mu} + 
    F_2^j(Q^2) {i \sigma^{\mu \nu} q_{\nu} \over 2 m} \right]
    \label{eq:one-body}
 \ee
where $Q^2 = - q \cdot q$. The approach of Ref.~\cite{nucleon} is characterized by the choice of a frame where $q^+ = 0$, which allows to eliminate the contribution of the so-called $Z$-graph (i.e., the pair creation from the vacuum \cite{Zgraph}). The important connection with the Feynmann triangle diagram is fully discussed in Ref.~\cite{MS} and the superiority of the choice $q^+ = 0$ for the one-body approximation (\ref{eq:one-body}) is clearly illustrated in Ref.~\cite{SIM02}.

\indent The matrix elements of the (on-shell) nucleon $e.m.$ current read as
 \be
    I_{\nu'_N \nu_N}^{\mu} & \equiv & \langle \Psi_N^{\nu'_N}| ~ J^{\mu} ~ 
    |\Psi_N^{\nu_N} \rangle \nonumber \\[2mm]
    & = & \bar{u}(P',\nu'_N) \left \{ F_1^N(Q^2) \gamma^{\mu} + F_2^N(Q^2) 
    {i \sigma^{\mu \nu} q_{\nu} \over 2M} \right \} u(P, \nu_N) ~~~~
    \label{eq:F1F2}
 \ee
where $u(P, \nu_N)$ is a Dirac spinor, $q = P' - P$ and $\Psi_N^{\nu_N}$ is the $LF$ wave function of the nucleon described in the Appendix, i.e. the same wave function used to calculate the momentum distribution $\overline{f}^N(z)$. In what follows we adopt a Breit frame where the four-momentum transfer $q \equiv (q^0, \vec{q})$ is given by $q^0 = 0$ and $\vec{q} = (q_x, q_y, q_z) = (Q, 0, 0)$. 

\indent The nucleon Sachs form factors are then given explicitly by \cite{nucleon}
 \be
    G_E^N(Q^2) = F_1^N(Q^2) - {Q^2 \over 4M} F_2^N(Q^2) = {1 \over 2} 
    \mbox{Tr}\{ I^+ [1 - {Q \over 2M} i\sigma_y] \}
    \label{eq:GEN}
 \ee
 \be
    G_M^N(Q^2) = F_1^N(Q^2) + F_2^N(Q^2) = - {P^+ \over Q} \mbox{Tr}\{ I^y i 
    \sigma_z \}
    \label{eq:GMN}
 \ee
where $\sigma_y$ and $\sigma_z$ are ordinary $2 \times 2$ Pauli matrices.

\begin{figure}[t]

\centerline{\epsfig{file=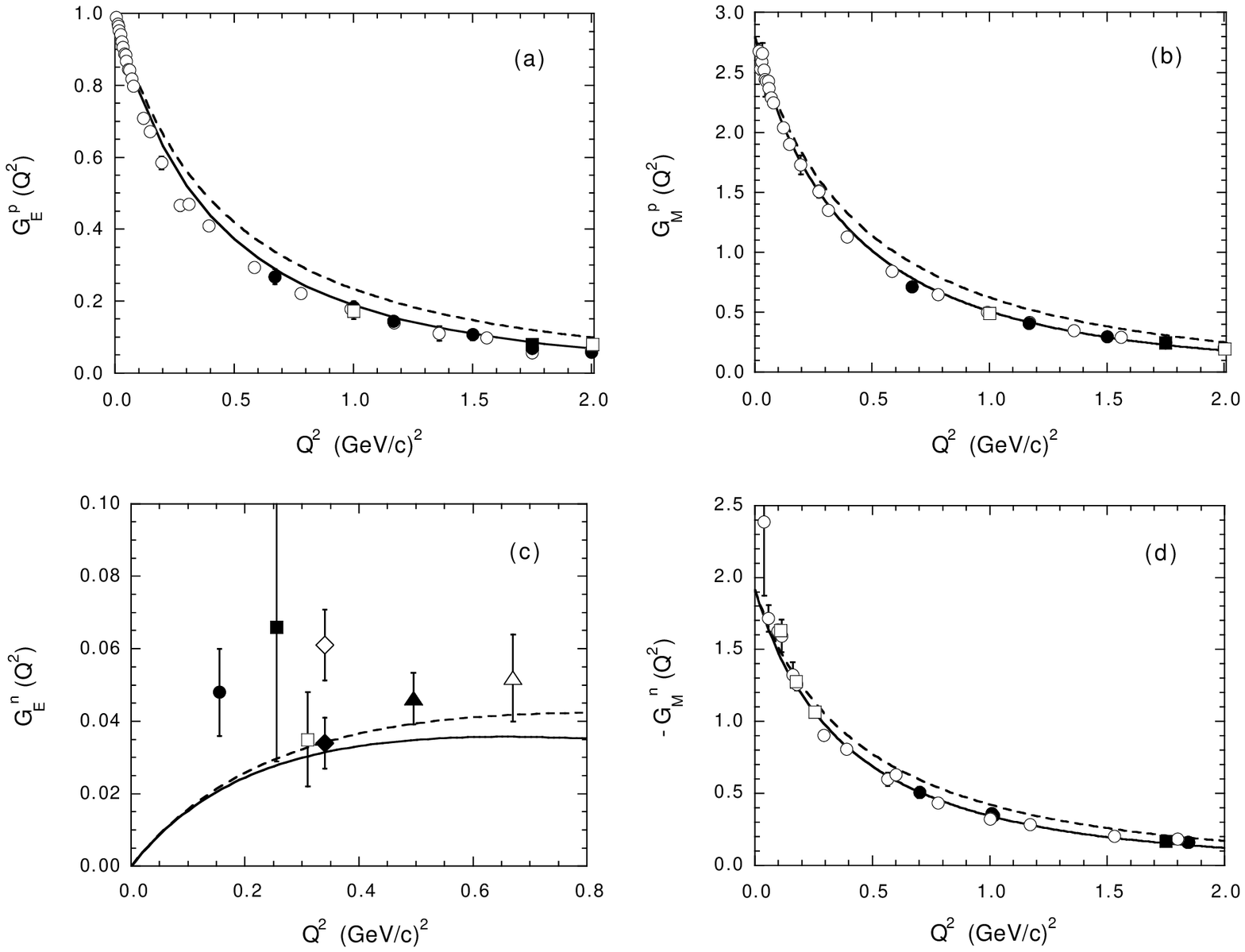,width=15.5cm}}

\caption{\label{fig:GiN} \small \em Elastic Sachs form factors of the nucleon, $G_E^p(Q^2)$ (a), $G_M^p(Q^2)$ (b), $G_E^n(Q^2)$ (c), $-G_M^n(Q^2)$ (d),  calculated using the covariant $LF$ approach of Ref.~\cite{nucleon}. The gaussian ans\"atz (\ref{eq:gaussian}) is adopted for the proton wave function with $\beta = 0.3 ~ GeV$ and $m = 0.25 ~ GeV$. In the one-body current (\ref{eq:one-body}) both Dirac and Pauli $CQ$ form factors are included, namely: $F_1^j(Q^2) = e_j / (1 + r_Q^2 ~ Q^2 / 6)$ and $F_2^j(Q^2) = \kappa_j / (1 + r_Q^2 ~ Q^2 / 12)^2$. The dashed and solid lines correspond to $r_Q = 0.21$ and $0.33 ~ fm$, respectively (see text). The values of the $CQ$ anomalous magnetic moments, $\kappa_U = - 0.064$ and $\kappa_D = 0.017$, have been fixed by the requirement of reproducing the experimental values of proton and neutron magnetic moments. In (a) and (b) full and open dots, open and full squares are the experimental data from Ref.~\cite{GEpGMp}(a-d), respectively. In (c) open squares, full squares, open diamonds, open triangles, full dots, full diamonds and triangles are the data from Ref.~\cite{GEn}(a-g), respectively. In (d) full dots, open dots, full and open squares are the data from Ref.~\cite{GMn}(a-d), respectively.}

\end{figure}

\indent We have then calculated Eqs.~(\ref{eq:GEN}-\ref{eq:GMN}) using the gaussian ans\"atz (\ref{eq:gaussian}) for the nucleon wave function and adopting the one-body approximation (\ref{eq:one-body}) with both Dirac and Pauli $CQ$ form factors having the following simple behavior: $F_1^j(Q^2) = e_j / (1 + r_Q^2 ~ Q^2 / 6)$ and $F_2^j(Q^2) = \kappa_j / (1 + r_Q^2 ~ Q^2 / 12)^2$. The values of the $CQ$ anomalous magnetic moments, $\kappa_U$ and $\kappa_D$, are fixed by the requirement of reproducing the experimental values of proton and neutron magnetic moments. The results of the calculations performed with the same parameters adopted in case of the ratio $R_n^p(Q^2)$ shown in Fig.~\ref{fig:threshold}, namely $\beta = 0.3 ~ GeV$, $m = 0.25 ~ GeV$ and $r_Q = 0.21 ~ fm$, are reported in Fig.~\ref{fig:GiN} as the dashed lines. Note that the combination $[F(Q^2)]^2$ given by Eq.~(\ref{eq:FCQ}), which is the one relevant for the scaling function (\ref{eq:scaling}), turns out to be almost totally dominated by the contribution of the Dirac form factors $F_1^j(Q^2)$ and it is basically insensitive to the presence of the Pauli form factors $F_2^j(Q^2)$. 

\indent It can be seen that the calculated form factors slightly overestimate the data, so that we can conclude that as a first approximation the scaling function of Fig.~\ref{fig:threshold} may be interpreted as a squared $CQ$ form factor. A better consistency with the data can be reached through slight variations of the parameters of our model, namely $r_Q$ and $\beta / m$. For instance, a nice agreement with the elastic data can be simply recovered by increasing the $CQ$ size up to $r_Q = 0.33 ~ fm$, as shown by the solid lines in Fig.~\ref{fig:GiN}. However, we can also ascribe the origin of the small discrepancies with the elastic data to the fact that the effects of the dynamical correlations among the $CQ$'s in the final state are so far missing in our low-$Q^2$ model. As already pointed out, the inclusion of such effects is not an easy task and it is well beyond the aim of the present paper.

\section{Conclusions}

\indent In this work we have first generalized the two-stage model of Refs.~\cite{Altarelli,APR}, originally developed in the $DIS$ regime, to values of $Q^2$ below the scale of chiral symmetry breaking and above the $QCD$ confinement scale, i.e. $0.1 \div 0.2 \lsim Q^2 ~ (GeV/c)^2 \lsim 1 \div 2$. The essential ingredient is the inclusion of the contribution to the inelastic hadronic structure functions arising from the {\em elastic} coupling at the constituent quark level. We have shown that within such a model a new scaling property [see Eq.~(\ref{eq:scaling})] is expected to occur in the inelastic hadronic structure functions, provided a reasonable model for the wave function describing the motion of the constituents inside the hadron is considered. Moreover, the resulting scaling function can be interpreted as the (squared) form factor of the constituent quark, i.e. the form factor of a confined object.

\indent Then we have analyzed the recent experimental determinations of the Nachtmann moments of the inelastic structure function of the proton $F_2^p(x, Q^2)$, obtained at $JLab$ \cite{JLAB}, for values of $Q^2$ ranging from $\approx 0.1$ to $\approx 2 ~ (GeV/c)^2$. The important results we have obtained are:

\begin{itemize}

\item{the scaling property (\ref{eq:scaling}) is well satisfied by the data;}

\item{the $CQ$ form factor extracted from the {\em inelastic} proton data is overall consistent with the one required to explain the {\em elastic} nucleon data;}

\item{the constituent quark size turns out to be $\approx 0.2 \div 0.3 ~ fm$.}

\end{itemize}

\indent Our findings clearly suggest that at low momentum transfer the inclusive proton structure function $F_2^p(x, Q^2)$ originates mainly from the elastic coupling with {\em extended objects inside the proton}. 

\indent A crucial, mandatory check of the extracted constituent form factor is provided by the analysis of the moments of the polarized proton structure function $g_1^p(x, Q^2)$. Indeed for $0.1 \div 0.2 \lsim Q^2  ~ (GeV/c)^2 \lsim 1 \div 2$ a scaling property analogous to Eq.~(\ref{eq:scaling}) is expected to hold also for the Nachtmann moments of $g_1^p(x, Q^2)$. The crucial point is that the two scaling functions, corresponding to the non-polarized and polarized cases, should coincide and provide the same constituent quark form factor.

\indent Measurements of $g_1^p(x, Q^2)$ at low values of $Q^2$ are still undergoing at $JLab$.

\newpage

\section*{Appendix: The nucleon light-front wave function}

\indent In this Appendix we briefly recall the basic notations and the relevant structure of the nucleon wave function in the Hamiltonian $LF$ formalism (see \cite{LF}). The nucleon $LF$ wave function is eigenstate of the non-interacting $LF$ angular momentum operators $j^2$ and $j_z$, where the unit vector $\hat{z} = (0, 0, 1)$ defines the spin quantization axis. The squared free-mass operator is given by
 \be
    M_0^2 = \sum_{i = 1}^3 ~ {|\vec{k}_{i \perp}|^2 + m^2 \over \xi_i}
    \label{eq:M0_LF}
 \ee
where $m$ is the mass of the constituent $U$ and $D$ quarks and
 \be
    \xi_i & = & {p_i^+ \over P^+} ~ , \nonumber \\
    \vec{k}_{i \perp} & = & \vec{p}_{i \perp} - \xi_i \vec{P}_{\perp}
    \label{eq:LF_var}
 \ee
are the intrinsic $LF$ variables. The subscript $\perp$ indicates the projection perpendicular to the spin quantization axis and the {\em plus} component of a 4-vector $p \equiv (p^0, \vec{p})$ is given by $p^+ = p^0 + \hat{z} \cdot \vec{p}$; finally $\tilde{P} \equiv (P^+, \vec{P}_{\perp}) = \tilde{p}_1 + \tilde{p}_2 + \tilde{p}_3$ is the nucleon $LF$ momentum and $\tilde{p}_i$ the $CQ$ one. Note that $\sum_i \xi_i = 1$.

\indent In terms of the longitudinal momentum $k_{iz}$, related to the variable $\xi_i$ by
 \be
    k_{iz} \equiv {1 \over 2} \left( \xi_i M_0 - { |\vec{k}_{i \perp}|^2 + 
    m^2 \over \xi_i M_0} \right)
    \label{eq:kin}
 \ee
the free mass operator acquires a familiar form, viz. 
 \be
     M_0 = \sum_{i = 1}^3 E_i =  \sum_{i = 1}^3 \sqrt{m^2 + |\vec{k}_i|^2}
    \label{eq:M0}
 \ee
with the three-vectors $\vec{k}_i$ defined as
 \be
   \vec{k}_i \equiv ( \vec{k}_{i \perp}, k_{iz})
   \label{eq:ki}
 \ee
Note that $\vec{k}_i$ are internal variables satisfying $\vec{k}_1 + \vec{k}_2 + \vec{k}_3 = 0$. Disregarding the color variables, the nucleon $LF$ wave function reads as
 \be
    \langle \{ \xi_i \vec{k}_{i \perp}; \nu'_i \tau_i \}| \Psi_N^{\nu_N} 
    \rangle = \sqrt{{E_1 E_2 E_3 \over M_0 \xi_1 \xi_2 \xi_3}} \sum_{ 
    \{\nu_i \}} ~ \langle \{ \nu'_i \} | {\cal{R}}^{\dag}|\{\nu_i \} \rangle 
    \cdot \langle \{\vec{k_i}; \nu_i \tau_i \} | \chi_N^{\nu_N} \rangle 
    \label{eq:wfLF} 
 \ee 
where $\nu_N$ is the third component of the nucleon spin, the curly braces $\{ ~~ \}$ mean a list of indexes corresponding to $i = 1, 2, 3$, and $\nu_i$ ($\tau_i$) is the third component of the $CQ$ spin (isospin). The rotation  ${\cal{R}}^{\dag}$, appearing in Eq.~(\ref{eq:wfLF}), is the product of individual (generalized) Melosh  rotations, viz.
 \be
     {\cal{R}}^{\dag} = \prod_{j = 1}^3  R_j^{\dag}(\vec{k}_{j \perp}, 
     \xi_j, m)
     \label{eq:Rmelosh}
 \ee
where \cite{Melosh}
 \be
    R_j(\vec{k}_{j \perp}, \xi_j, m) \equiv {m + \xi_j M_0 - i 
    \vec{\sigma}^{(j)} \cdot (\hat{n} \times \vec{k}_{j \perp}) \over 
    \sqrt{(m + \xi_j M_0)^2 + |\vec{k}_{j \perp}|^2}}
    \label{eq:melosh}
 \ee
with $\vec{\sigma}$ being the ordinary Pauli spin matrices.

\indent Neglecting the very small $P$- and $D$-waves in the nucleon (cf. \cite{nucleon}) we can limit ourselves to canonical (or equal-time) wave function corresponding to a total orbital angular momentum equal to $L = 0$; one has
 \be
    \langle \{ \vec{k}_i; \nu_i \tau_i\}| \chi_N^{\nu_N} \rangle & = & 
    w_S(\vec{k}, \vec{p}) {1 \over \sqrt{2}} \left[ \Phi^{00}_{\nu_N \tau_N} 
    + \Phi^{11}_{\nu_N \tau_N} \right] \nonumber \\
    & + & w_{S'_s}(\vec{k}, \vec{p}) {1 \over \sqrt{2}} \left[ 
    \Phi^{00}_{\nu_N \tau_N} - \Phi^{11}_{\nu_N \tau_N} \right] \nonumber \\
    & + & w_{S'_a}(\vec{k}, \vec{p}) {1 \over \sqrt{2}} \left[ 
    \Phi^{01}_{\nu_N \tau_N} + \Phi^{10}_{\nu_N \tau_N} \right] \nonumber \\
    & + & w_{A}(\vec{k}, \vec{p}) {1 \over \sqrt{2}} \left[ 
    \Phi^{01}_{\nu_N \tau_N} - \Phi^{10}_{\nu_N \tau_N} \right] ~~~~
    \label{eq:canonical}
 \ee
where $w_S(\vec{k}, \vec{p})$, $w_{S'_s}(\vec{k}, \vec{p})$, $w_{S'_a}(\vec{k}, \vec{p})$ and $w_A(\vec{k}, \vec{p})$ are the completely symmetric ($S$), the two mixed-symmetry ($S'_s$ and $S'_a$) and the completely antisymmetric ($A$) wave functions, respectively. In Eq.~(\ref{eq:canonical}) the variables $\vec{k}$ and $\vec{p}$ are the Jacobian internal coordinates, defined as
 \be
   \vec{k} & = & {\vec{k}_1 - \vec{k}_2 \over 2} ~ , 
   \nonumber \\
   \vec{p} & = & {2 \vec{k}_3 - \left( \vec{k}_1 + \vec{k}_2 \right) \over 
    3}
   \label{eq:jacobian}
 \ee
with $\vec{k}_i$ given by Eq.~(\ref{eq:ki}). Finally, the spin-isospin function $\Phi^{S_{12} T_{12}}_{\nu_N \tau_N}$, corresponding to a total spin $(1/2)$ and total isospin $(1/2)$, is defined as
 \be
    \Phi^{S_{12} T_{12}}_{\nu_N\tau_N} & = & \sum_{M_S} \langle {1 \over 2} 
    \nu_1 {1 \over 2} \nu_2 | S_{12} M_S \rangle ~ \langle S_{12} M_S {1 
    \over 2} \nu_3 | {1 \over 2} \nu_N \rangle \nonumber \\
    & \cdot & \sum_{M_T} \langle {1 \over 2} \tau_1 {1 \over 2} \tau_2 | 
    T_{12} M_T \rangle ~ \langle T_{12} M_T {1 \over 2} \tau_3 | {1 \over 2} 
    \tau_N \rangle ~~~~~~~~
    \label{eq:STwf}
 \ee
where $S_{12}$ ($T_{12}$) is the total spin (isospin) of the quark pair $(1, 2)$. The normalization of the various partial waves in Eq.~(\ref{eq:canonical}) is
 \be
    \int d\vec{k} d\vec{p} \left| w_S(\vec{k}, \vec{p}) \right|^2 & = & 
    P_S ~ , \nonumber \\
    \int d\vec{k} d\vec{p} \left| w_{S'_s}(\vec{k}, \vec{p}) \right|^2 & = 
    & \int d\vec{k} d\vec{p} \left| w_{S'_a}(\vec{k}, \vec{p}) \right|^2 = 
    P_{S'} / 2 ~ , \nonumber \\
    \int d\vec{k} d\vec{p} \left| w_A(\vec{k}, \vec{p}) \right|^2 & = & 
    P_A ~ ,
    \label{eq:norms}
 \ee
with $P_S + P_{S'} + P_A = 1$.

\indent Disregarding the completely antisymmetric component $w_A(\vec{k}, \vec{p})$, which is usually quite negligible in the nucleon (cf. \cite{nucleon}), the constituent $U$ and $D$ momentum distributions, defined in Eq.~(\ref{eq:fQz}), read explicitly as
 \be
     f_U^p(z) & = & 2 ~ \int d\vec{k}_{\perp} d\vec{p}_{\perp} \int [d\xi_i] 
     ~ \delta(z - \xi_1) ~ {E_1 E_2 E_3 \over M_0 \xi_1 \xi_2 \xi_3} 
     \nonumber \\
     & \cdot & \left[ |w_S(\vec{k}, \vec{p})|^2 + |w_{S'_s}(\vec{k}, 
     \vec{p})|^2 + |w_{S'_a}(\vec{k}, \vec{p})|^2 + w_S(\vec{k}, \vec{p}) ~ 
     w_{S'_s}(\vec{k}, \vec{p}) \right] ~, \nonumber \\[4mm]
     f_D^p(z) & = & \int d\vec{k}_{\perp} d\vec{p}_{\perp} \int [d\xi_i] 
     ~ \delta(z - \xi_1) ~ {E_1 E_2 E_3 \over M_0 \xi_1 \xi_2 \xi_3} 
     \nonumber \\
     & \cdot & \left[ |w_S(\vec{k}, \vec{p})|^2 + |w_{S'_s}(\vec{k}, 
     \vec{p})|^2 + |w_{S'_a}(\vec{k}, \vec{p})|^2 - 2 w_S(\vec{k}, \vec{p}) 
     ~ w_{S'_s}(\vec{k}, \vec{p}) \right]
     \label{eq:fUfD}
 \ee
It can be seen that the relativistic composition of the $CQ$ spins (i.e. the Melosh rotations) do not affect at all the (unpolarized) $LF$ momentum distribution $f_Q^p(z)$. Moreover, any flavor dependence of $f_Q^p(z)$ turns out to be driven by the interference between the completely symmetric ($S$) and mixed-symmetry ($S'_s$) wave functions, the latter being generated mainly by the spin-spin component of the interaction among $CQ$'s which are present both in the one-gluon-exchange model of Ref.~\cite{CI} and in the chiral model of Ref.~\cite{Glozmann}. In the limit of exact $SU(6)$ symmetry one has $w_{S'_s} = w_{S'_a} = 0$ and Eq.~(\ref{eq:SU6}) is recovered.

\end{document}